\documentclass{IEEEtran}
\usepackage{graphicx}
\usepackage{bm}
\usepackage{cite}
\usepackage{pgfplots}
\usepackage{color}
\usepackage{algorithm}
\usepackage{amsmath,amssymb,amsthm,dsfont,mathrsfs}

\def\diag{{\mathrm{diag}}}
\def\tr{{\mathrm{tr}}}
\def\A{{\mathbf{A}}}
\def\Y{{\mathbf{Y}}}
\def\S{{\mathbf{S}}}
\def\I{{\mathbf{I}}}
\def\N{{\mathbf{N}}}
\def\B{{\mathbf{B}}}
\def\y{{\mathbf{y}}}
\def\s{{\mathbf{s}}}
\def\a{{\mathbf{a}}}
\def\n{{\mathbf{n}}}
\def\b{{\mathbf{b}}}

\begin{document}
%
\title{Root Sparse Bayesian Learning for Off-Grid DOA Estimation
}

\author{
\authorblockN{Jisheng~Dai,~\IEEEmembership{Member,~IEEE},\thanks{J. Dai is with the Department of Electronic  Engineering, Jiangsu University, Zhenjiang 212013, China, and the National Mobile Communications Research Laboratory, Southeast University, Nanjing 210096, China (e-mail: jsdai@ujs.edu.cn).}}
\authorblockN{Xu~Bao, \thanks{X. Bao is with the Department of Telecommunication Engineering, Jiangsu University, Zhenjiang 212013, China (e-mail: xbao@ujs.edu.cn).}}
\authorblockN{Weichao~Xu,~\IEEEmembership{Member,~IEEE},\thanks{W. Xu is with the Department of Automatic Control, Guangdong University of Technology, Guangzhou 510006, China (E-mail: wcxu@gdut.edu.cn).}}
and~\authorblockN{Chunqi~Chang,~\IEEEmembership{Member,~IEEE}\thanks{C. Chang is with the School of Biomedical Engineering, Shenzhen University, Shenzhen 518060, China (e-mail: cqchang@szu.edu.cn).}}
\thanks{Copyright $\copyright$ 2016 IEEE. Personal use of this material is permitted. However, permission to use this material for any other purposes must be obtained from the IEEE by sending a request to pubs-permissions@ieee.org.}
}

\maketitle

\begin{abstract}
The performance of the existing  sparse Bayesian learning (SBL) methods for off-grid DOA estimation is dependent on the trade off  between the accuracy and the computational workload.
To speed up the off-grid SBL method while remain a reasonable accuracy,
this letter describes a computationally efficient root SBL method for off-grid DOA estimation,
which adopts a coarse grid and
consider the sampled locations in the coarse grid as
the adjustable parameters.
We utilize
an expectation-maximization (EM) algorithm to iteratively
refine this coarse grid, and illustrate that each  updated
grid point can be simply achieved by the root of a certain
polynomial.  Simulation results demonstrate that the computational complexity is
significantly reduced and the modeling error can be almost
eliminated.
\end{abstract}


\IEEEpeerreviewmaketitle

\section{Introduction}

Direction-of-arrival (DOA) estimation of narrow-band sources is an important topic in array signal processing and has attracted tremendous interest in many fields, such as radar, sonar, mobile communications \cite{krim1996two}.
The conventional subspace-type DOA estimation methods, including MUSIC and ESPRIT, require a large number of snapshots to
achieve their high-resolution performance, and they may fail to work when signals are highly correlated or coherent due to multipath propagation. Recently, the emerging technique of sparse representation has given renewed interest to the DOA estimation problem \cite{malioutov2005sparse,carlin2013directions,huang2016real,dai2013direction}. The sparse representation methods exhibit many advantages compared to the conventional subspace-type ones, e.g., improved robustness
to noise, limited number of snapshots and correlation of signals.

Sparse Bayesian learning (SBL) \cite{tipping2001sparse,zhao2013improved} is one of the most popular approaches for the sparse signal recovery, which exploits
the sparsity information with a sparse prior assumption for the signal of interest from a Bayesian perspective.
Theoretical results show that SBL includes the $l_1$-norm minimization method as a special case when a maximum $a$ $posteriori$ (MAP) optimal estimate is
adopted with a Laplace signal prior \cite{ji2008bayesian,wipf2004sparse}. The excellent performance achieved by the SBL methods for DOA estimation relies crucially on the assumption that the true DOAs lie on (practically, close to) the sampling grid points. In practice, however, this assumption is usually unavailable. The gap between the true DOA and its nearest grid point is known as the off-grid gap.


A number of improved methods have been proposed to deal with the off-grid DOA estimation \cite{zhu2011sparsity,yang2013off,zhang2014off}.
Yang et al. \cite{yang2013off} applied a linear approximation to the true DOA and proposed an off-grid SBL method, where the off-grid gap is assumed to be uniformly distributed (noninformative). Using the sample covariance matrix, Zhang et al. \cite{zhang2014off} further gave an improved off-grid SBL method to reduce the effect of noise variance.
The existing methods \cite{yang2013off,zhang2014off}  remain a major problem that their performance is dependent on the trade off between the accuracy and the computational workload.
When a coarse grid is used, it may lead to a high modeling error; while,
if a dense sampling grid is used, the massive involved computation could make this method unfavorable for real applications.

In this letter, we try to propose a computationally efficient root SBL method for off-grid DOA estimation,
which adopts a coarse grid and
considers the sampled locations in the coarse grid as
the adjustable parameters.
We utilize  an expectation-maximization (EM) algorithm to iteratively refine this coarse grid, and illustrate that each  updated grid point can be simply achieved by the root of a certain polynomial. In this way, the computational complexity is significantly reduced and the modeling error can be almost eliminated.


%

\section{Data Model}
Consider $K$ narrow-band far-field sources impinging on an $M$-element
uniform linear array (ULA), where the distance between adjacent sensors is $d$. The $K$ sources, $s_1(t), s_2(t),\ldots, s_K(t)$, arrive at the array from distinct directions,
$\theta_1,\theta_2,\ldots,\theta_K$, with respect to the normal line
of the array. The $M\times 1$ array output vector $\y (t)$ is then given by
\begin{eqnarray}\label{model1}
\y(t)=\A\s(t)+\n(t),~~~t\in\{t_1,t_2,\ldots,t_T\}
\end{eqnarray}
where $\mathbf{y}(t)=[y_1(t),y_2(t),\ldots,y_M(t)]^T$, $\mathbf{s}(t)=[s_1(t),s_2(t),
\ldots,s_K(t)]^T$,
${\A}=[{\bm{\alpha}}(\theta_1), {\bm{\alpha}}(\theta_2), \ldots,{\bm{\alpha}}(\theta_K)]$,
$\bm{\alpha}(\theta_k)=[
1, v_{\theta_k}, \ldots, v_{\theta_k}^{M-1}]^T$, $v_{\theta_k}=e^{ -j2\pi d/\lambda \sin (\theta_k)}$,
$\lambda$ is the wavelength of the source, and
$\mathbf{n}(t)=\left[n_1(t),n_2(t),\ldots,n_M(t)\right]^T$ is an unknown noise vector.
With the definitions of $\Y\triangleq[\y(t_1),\y(t_2),\ldots,\y(t_T)]$, $\N\triangleq[\n(t_1),\n(t_2),\ldots,\n(t_T)]$ and $\S\triangleq[\s(t_1),\s(t_2),\ldots,\s(t_T)]$, we have
\begin{align}\label{eq:R}
\Y= {\A}\S + \N.
\end{align}
In order to cast the problem of DOA estimation as a sparse representation problem, generally we let $\{\hat\theta_i\}_{i=1}^{\hat{K}}$ be a fixed sampling grid that uniformly covers the DOA range $[-\pi/2, \pi/2]$, where $\hat{K}$ denotes the grid number. If the grid is fine enough such that the true DOAs lie on (or, practically, close to) the grid, we can use the following model for $\Y$:
\begin{eqnarray}\label{problem1}
\Y = {\A}_{\hat\theta} \hat{\S}+ \N
\end{eqnarray}
where $\A_{\hat\theta}\triangleq [ {\bm\alpha} (\hat{\theta}_1),  \ldots,  {\bm\alpha} (\hat{\theta}_{\hat{K}}) ]$ and $\hat{\S}$ is a $\hat{K}\times T$ complex matrix whose the $i$th row corresponds to the signal impinging on the array from a possible source at $\hat\theta_i$. However, the assumption that the true DOAs are located on the predefined spatial grid is not always valid in practical implementations.

To solve this problem, a  linear approximation method is used in \cite{yang2013off}. If $\theta_k\notin \{\hat\theta_i\}_{i=1}^{\hat{K}}$ and $\hat\theta_{n_k}, n_k\in\{1,2,\ldots, \hat{K}\}$, is the nearest grid point to $\theta_k$, the steering vector $\bm\alpha(\theta_k)$ is approximated by the linearization:
\begin{align}
\bm\alpha(\theta_k)\approx \bm\alpha(\hat\theta_{n_k}) + \b(\hat\theta_{n_k}) \left(\theta_k-  \hat\theta_{n_k}  \right)
\end{align}
where $\b(\hat\theta_{n_k}) =  \bm\alpha'(\hat\theta_{n_k})$.
By absorbing the approximation error into the noise, the observation model (\ref{problem1}) can be rewritten as
\begin{eqnarray}\label{problem2}
\Y = \bm\Phi(\bm\zeta) \hat{\S}   +   \N
\end{eqnarray}
where $\bm\Phi(\bm\zeta)=  {\A}_{\hat\theta} + \B \diag\{ \bm\zeta \}$, $\B=[\b(\hat\theta_{1}),\ldots, \b(\hat\theta_{\hat{K}})]$, $ \diag\{ \cdot\}$ represents the diagonal matrix operator,
and $\bm\zeta$ is a zero vector except that the $n_k$-th element
$[\bm\zeta]_{n_k}= \theta_k - \hat\theta_{n_k}, k=1,2,\ldots, K.$


Obviously, the modeling error caused by off-grid gap can be alleviated by the new model (\ref{problem2}), but it can not be fully eliminated. When a coarse grid is used, (\ref{problem2}) may still lead to a high modeling error; while,
if a dense sampling grid is used, the massive involved computation could make the linear approximation method unfavorable for real applications.

\section{The Proposed Root SBL Method}

In this section, we try to propose a computationally efficient root SBL method for off-grid DOA estimation. To keep a low computational load, we adopt the original model (\ref{problem1}) with a coarse grid (i.e., $\hat K$ is small).
To handle the modeling error in (\ref{problem1}), we consider the sampled locations in the coarse grid as the adjustable parameters, and utilize an expectation-maximization (EM) algorithm to iteratively refine the grid. 
Note that the name is due to the fact that each
updated grid point can be simply achieved by using the root of
a polynomial.

\subsection{Sparse Bayesian Formulation}
Firstly, we address the sparse Bayesian model that is commonly used in SBL \cite{tipping2001sparse}. A typical SBL treatment of $\hat{\S}$ begins by assigning a non-stationary Gaussian prior distribution with a distinct inverse variance $\delta_i$ for each row of $\hat{\S}$.
Letting $\bm{\delta} =[\delta_1, \delta_2,\ldots, \delta_{\hat{K}}]^T$ and $\bm{\Delta}= \mathrm{diag}(\bm{\delta} )$, we have
\begin{align}
p(\hat{\S}|\bm{\delta})= \prod_{t=1}^{T} \mathcal{CN}(\hat\s_t| \bm{0},  \bm{\Delta} )\label{eq:mos}
\end{align}
where $\hat\s_t$ denotes the $t$th column of $\hat{\S}$.
In order to obtain a two-stage hierarchial prior that favors most rows of $\hat{\S}$ being zeros, the hyper-parameter $\delta_i$s are further modeled as independent Gamma distributions \cite{tipping2001sparse,ji2008bayesian}, i.e.,
\begin{align}
p(\bm{\delta})= \prod_{i=1}^{\hat{K}}  \Gamma(\delta_i;~1,\rho )
\end{align}
where $\rho$ is a small positive constraint (e.g., $\rho=0.01$ \cite{ji2008bayesian}).

Under an assumption of circular symmetric complex Gaussian noises,
we have
\begin{align}
p(\Y | \hat{\S}, \beta; \hat{\bm \theta}) = \prod_{t=1}^{T} \mathcal{CN}(\y_t |{\A}_{\hat\theta} \hat\s_t, \beta^{-1}\I)
\end{align}
where $\beta\triangleq\sigma^{-2}$ denotes the noise precision with $\sigma^{2}$ being the noise variance, $\hat{\bm \theta}\triangleq[ \hat{\theta}_1, \hat{\theta}_2 \ldots,  \hat{\theta}_{\hat{K}} ]$ and $\y_t$ denotes the $t$th column of $\Y$.
As $\beta$ is usually unknown, we model it as a Gamma hyperprior
$p(\beta)=\Gamma(\beta;~a,b)$,
where we set $a,b\rightarrow 0$ as in \cite{tipping2001sparse,ji2008bayesian} so as to obtain a broad hyperprior.

\subsection{Bayesian Inference}
As $p(\hat{\S}, \beta,\bm{\delta}|\Y;\hat{\bm \theta} )$ cannot be explicitly calculated,
an EM algorithm will be exploited  to perform the Bayesian inference. The principle behind the EM algorithm is to
repeatedly construct a lower-bound on the evidence function $p(\beta,\bm{\delta} |\Y; \hat{\bm \theta})$, or equivalently, $\ln p(\Y, \beta,\bm{\delta}; \hat{\bm \theta})$ (E-step), and then optimize that lower-bound (M-step).

In E-step, we treat $\hat\S$ as a hidden variable, whose  posterior distribution is  also a complex Gaussian \cite{tipping2001sparse}:
\begin{align}\label{eq:gau}
p(\hat\S|\Y, \beta,\bm{\delta};~ \hat{\bm \theta})
=\prod_{t=1}^{T} \mathcal{CN}( \hat\s_t | \bm\mu_t, \bm\Sigma)
\end{align}
where
\begin{align}
\bm\mu_t &= \beta \bm\Sigma {\A}_{\hat\theta}^H \y_t,~~~t=1,2,\ldots, T\\
\bm\Sigma &=  ( \beta {\A}_{\hat\theta}^H {\A}_{\hat\theta}  + \bm\Delta^{-1}      )^{-1}.
\end{align}
The well known lower-bound on $\ln p(\Y, \beta,\bm{\delta}; \hat{\bm \theta})$ is given by
\begin{align}
&\mathcal{L}(\beta,\bm{\delta}; \hat{\bm \theta})= \left<\ln~  p(\hat{\S},\Y, \beta,\bm{\delta};\hat{\bm \theta}  ) \right>_{p(\hat\S|\Y, \beta,\bm{\delta};\hat{\bm \theta})}\\
=& \left<\ln~  p(\Y | \hat\S, \beta; \hat{\bm \theta} ) p(\hat\S|\bm{\delta})
p(\beta) p(\bm{\delta}) \right>_{p(\hat\S|\Y, \beta,\bm{\delta};\hat{\bm \theta})}\label{eqbd2}
\end{align}
where (\ref{eqbd2}) aligns with the hierarchical Bayesian model.


In M-step,  we first give the hyperparameter updates for $\bm\delta$ and $\beta$:
\begin{align}
\delta_i^{new}&=\frac{- T  + \sqrt{  T^2+ 4\rho \sum_{t=1}^{ T} [\bm\Xi_t]_{ii} } }{2\rho}\label{eq-up1}\\
\beta^{new}&=\frac{{T}M + (a-1)}
{ b+\sum_{t=1}^{ T} \|\y_t- {\A}_{\hat\theta} \bm\mu_t  \|_2^2+ T\tr\left({\A}_{\hat\theta}\bm\Sigma {\A}_{\hat\theta}^H \right)
}.\label{eq-up2}
\end{align}
where $\bm\Xi_t\triangleq \bm\mu_t(\bm\mu_t)^H + \bm{\Sigma}$. As (\ref{eq-up1}) and (\ref{eq-up2}) can be obtained
by following the similar procedure as in \cite{tipping2001sparse} (also refer to \cite{yang2013off}), their derivations are omitted for brevity.

In the rest part, we will focus on the  parameter update for $\hat{\bm\theta}$. Ignoring terms in the logarithm independent thereof, we just have to maximize
\begin{align}
&\left<\ln~  p(\Y | \hat\S, \beta; \hat{\bm \theta} )  \right>_{p(\hat\S|\Y, \beta,\bm{\delta};\hat{\bm \theta})}\notag\\
=&  -\beta\sum_{t=1}^{T}  \left<
\|\y_t- {\A}_{\hat\theta} \hat\s_t  \|_2^2  \right>_{p(\hat\S|\Y, \beta,\bm{\delta};\hat{\bm \theta})}   \\
=& -\beta\sum_{t=1}^{T} \|\y_t- {\A}_{\hat\theta} \bm\mu_t \|_2^2  -\beta T\tr\left( {\A}_{\hat\theta}\bm\Sigma {\A}_{\hat\theta}^H \right). \label{equptheta}
\end{align}
To refine each sampled location $\hat\theta_i, i=1,2,\ldots, \hat K $, or equivalently, its exponential form $v_{\hat{\theta}_i}(\triangleq e^{ -j2\pi d/\lambda \sin (\hat\theta_i)})$,  we calculate the derivative of (\ref{equptheta}), with respect to  $v_{\hat{\theta}_i}$, and then set it to zero:
\begin{align}\scriptsize
(\a_i')^H \left(\a_i\underbrace{\sum_{t=1}^T \left( |\mu_{ti}|^2 +\gamma_{ii} \right)}_{\triangleq \phi^{(i)} }  + \underbrace{T\sum_{j\ne i} \gamma_{ji}\a_j -  \sum_{t=1}^T\mu_{ti}^* \cdot \y_{t-i}}_{\triangleq \bm\varphi^{(i)}}\right)=0\label{eqrot}
\end{align}
where $\a_i$, $\mu_{ti}$ and $\gamma_{ij}$ denote the $i$th column, the $i$th element and the $(i,j)$th element of
$\A_{\hat\theta}$, $\bm\mu_t$ and  $\bm\Sigma$, respectively,
$\y_{t-i}\triangleq\y_t- \sum_{j\ne i} \mu_{tj} \a_j $, $\a_i'\triangleq {d \a_i}/{d v_{\hat{\theta}_i}} $, and $(\cdot)^*$ stands for conjugate operation.
Here, we use the fact that
\begin{align}
&\frac{\partial \sum_{t}\|\y_t- {\A}_{\hat\theta} \bm\mu_t \|_2^2}{\partial v_{\hat{\theta}_i} }
= (\a_i')^H  \left(\a_i\sum_{t=1}^T |\mu_{ti}|^2  -  \sum_{t=1}^T\mu_{ti}^* \cdot \y_{t-i}\right)\notag\\
&\frac{\partial \tr\left( {\A}_{\hat\theta}\bm\Sigma {\A}_{\hat\theta}^H \right) }{\partial v_{\hat{\theta}_i} }
=(\a_i')^H  \A_{\hat\theta} \bm\gamma_i=(\a_i')^H \left(\gamma_{ii}\a_i   + \sum_{j\ne i} \gamma_{ji}\a_j \right).\notag
\end{align}

After some algebraic operations, (\ref{eqrot}) can be rewritten in a polynomial form:
\begin{align}
[v_{\hat{\theta}_i},1,v_{\hat{\theta}_i}^{-1}, \ldots , v_{\hat{\theta}_i}^{-(M-2)} ]
\begin{bmatrix}
\frac{M(M-1)}{2}\phi^{(i)}\\
\varphi^{(i)}_2\\
2\varphi^{(i)}_3\\
\vdots\\
(M-1)\varphi^{(i)}_{M}
\end{bmatrix}=0\label{eqpolynomial}
\end{align}
where $\varphi^{(i)}_m\triangleq \left[\bm\varphi^{(i)}\right]_m$.
As the polynomial is of order $M-1$, it has $M-1$ roots in the complex plane.
According to the definition of $v_{\hat{\theta}_i}$, the selected root for refining $v_{\hat{\theta}_i}$ should be with a unit absolute value; however, due to the presence of noise, the roots may not be on the unit circle. In this case, the closest root to the unit circle is selected (which is denoted by $z_{i^\star}$), and the candidate point for the refined grid is
\begin{align}\label{eq-root}
\hat{\theta}_{i^\star}^{new}=\arcsin\left( -\frac{\lambda}{2\pi d}\cdot\mathrm{angle}(z_{i^\star}) \right).
\end{align}
On the other hand, we note that when the original coarse grid uniformly covers the DOA range, the DOA estimates on the rough grid is near the true DOAs. Hence, we further screen the candidate from where it falls into. $\hat{\theta}_{i^\star}^{new}$ is finally accepted if $\hat{\theta}_{i^\star}^{new}$ falls into the set of $\left[\frac{\hat{\theta}_{i^\star-1}+\hat{\theta}_{i^\star}}{2}, \frac{\hat{\theta}_{i^\star}+\hat{\theta}_{{i^\star}+1}}{2}\right]$; otherwise, it is rejected and the corresponding grid point retains unchanged.

Actually, we do not need to refine every $\hat\theta_i$ in each iteration, because
any $\hat\theta_i$s corresponding to rows that have small entries can be safely discarded.
In the practical implementation, we may set a  threshold to select some proper active grid points as follows. Let $f_t$ be the Frobenius norm of the mean of $\hat{\s}_t$.
Then, the indexes of the grid points that need to be activated can be selected by finding the first $\eta$ maxima $f_t$, where $1\le\eta\le M$.
Note that if we use a small $\eta$,  the selected grid points in the current iteration may miss some true DOAs. However, as the Frobenius norms $f_t$s will vary with the iteration, the missing DOAs might be activated in the next iterations. Simulation results recommend to set $\eta \ge K$ for grid-refining, especially $\eta = M$ if the number of sources $K$ is not available.

Since $\eta$ is much smaller than the grid number $\hat{K}$, the computation of (\ref{eq-root}) is negligible, and the most demanding steps for our method are in (\ref{eq-up1}) and (\ref{eq-up2}), whose computational complexity is of order $\mathcal{O}(M \hat{K}^2 )$ per iteration. It is the same as the OGSBI method \cite{yang2013off}. However, it is worth noting that our method can work with a very coarse grid, which will bring a significant improvement for computational cost.



\begin{figure}
\center
\begin{tikzpicture}[scale=0.75]
\begin{semilogyaxis}[xlabel={Grid interval [degree]},
ylabel={RMSE [degree]},grid=major,
legend style={at={(0.52,1.22),font=\footnotesize},
anchor=north,legend columns=2}, ]
\addplot[mark=asterisk,/tikz/dashed,red]  coordinates{
  (    1.0000,    0.6226)
  (    2.0000,    0.6680)
  (    4.0000,    0.8273)
  (    6.0000,    0.9872)
  (    8.0000,    1.0909)
  (   10.0000,    1.1392)
};
\addplot[mark=asterisk,red]  coordinates{
  (    1.0000,    0.1874)
  (    2.0000,    0.2286)
  (    4.0000,    0.2875)
  (    6.0000,    0.3075)
  (    8.0000,    0.3303)
  (   10.0000,    0.3616)
};

\addplot[mark=o,/tikz/dashed,blue]  coordinates{
  (    1.0000,    0.6547)
  (    2.0000,    0.7566)
  (    4.0000,    1.1747)
  (    6.0000,    1.6101)
  (    8.0000,    2.1589)
  (   10.0000,    2.5402)
};
\addplot[mark=o,blue] coordinates{
  (    1.0000,    0.2020)
  (    2.0000,    0.2970)
  (    4.0000,    0.4842)
  (    6.0000,    0.7422)
  (    8.0000,    1.2845)
  (   10.0000,    1.7003)
};

\addplot[mark=triangle,/tikz/dashed,black]  coordinates{
  (    1.0000,    0.6284)
  (    2.0000,    0.7710)
  (    4.0000,    1.2662)
  (    6.0000,    1.8722)
  (    8.0000,    2.5513)
  (   10.0000,    2.9938)
};
\addplot[mark=triangle,black]  coordinates{
  (    1.0000,    0.3318)
  (    2.0000,    0.5910)
  (    4.0000,    1.2134)
  (    6.0000,    1.8477)
  (    8.0000,    2.4580)
  (   10.0000,    2.9162)
};

\addplot[/tikz/dashed,black]  coordinates{
  (    1.0000,      0.5194)
  (    2.0000,      0.5179)
  (    4.0000,      0.5177)
  (    6.0000,      0.5188)
  (    8.0000,      0.5187)
  (   10.0000,      0.5174)
};
\addplot[black]  coordinates{
  (    1.0000,    0.1550)
  (    2.0000,    0.1556)
  (    4.0000,    0.1546)
  (    6.0000,    0.1546)
  (    8.0000,    0.1548)
  (   10.0000,    0.1551)
};
\legend{ {Our method, SNR=$0$dB},{Our method, SNR=$10$dB},{OGSBI, SNR=$0$dB~~~~~~}, {OGSBI, SNR=$10$dB~~~~~~}, {$l_1$-SVD, SNR=$0$dB~~~~~}, {$l_1$-SVD, SNR=$10$dB~~~~~}, {CRB, SNR=$0$dB~~~~~~~~~},{CRB, SNR=$10$dB~~~~~~~~}  }
\end{semilogyaxis}
\end{tikzpicture}
\caption{RMSE of DOA estimate versus grid interval.
 }\label{fig1}
\end{figure}
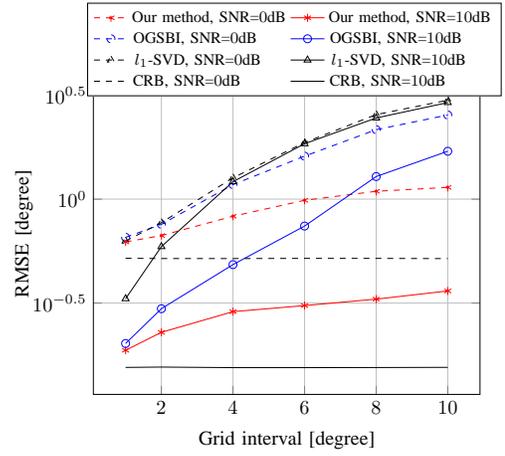

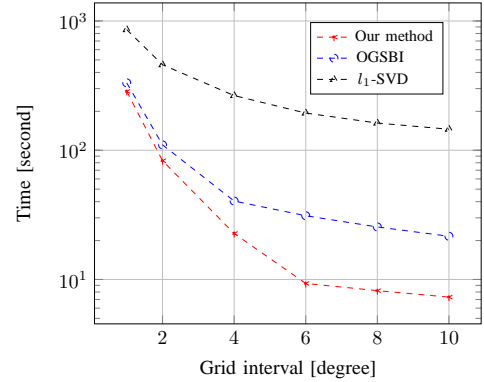
\begin{figure}
\center
\begin{tikzpicture}[scale=0.75]
\begin{semilogyaxis}[xlabel={Grid interval [degree]},
ylabel={Time [second]},grid=major,
legend style={at={(0.73,0.94),font=\footnotesize},
anchor=north,legend columns=1}
]
\addplot[mark=asterisk,red,/tikz/dashed]  coordinates{
  (    1.0000,  285.4428)
  (    2.0000,   82.9011)
  (    4.0000,   22.6931)
  (    6.0000,    9.3143)
  (    8.0000,    8.1966)
  (   10.0000,    7.3026)
};
\addplot[mark=o,blue,/tikz/dashed] coordinates{
  (    1.0000,  332.2810)
  (    2.0000,  110.0493)
  (    4.0000,   40.4461)
  (    6.0000,   31.1209)
  (    8.0000,   25.4964)
  (   10.0000,   21.5872)
};
\addplot[mark=triangle,black,/tikz/dashed]   coordinates{
  (    1.0000,  856.9930)
  (    2.0000,  458.5242)
  (    4.0000,  264.8359)
  (    6.0000,  193.9470)
  (    8.0000,  162.3489)
  (   10.0000,  145.3860)
};
\legend{ Our method, OGSBI~~~~~~, $l_1$-SVD~~~~ }
\end{semilogyaxis}
\end{tikzpicture}
\caption{Computational time versus grid interval with SNR $=$ $0$ dB.}
\label{fig2}
\end{figure}

\section{Simulation Results}
In this section, we will present several simulation results to illustrate the  performance of our proposed method. We will compare the proposed method to OGSBI \cite{yang2013off} and the $l_1$-SVD method in \cite{malioutov2005sparse}, as well as the Cramer-Rao bound (CRB).
The experiments are carried out in MATLAB 8.3.0 on a PC with an AMD FX(tm)-8350 CPU and 16GB of RAM.
Matlab codes have been made available online at https://sites.google.com/site/jsdaiustc/publication.

Simulation~1 verifies the performance improvement of the proposed method in terms of the root mean square error (RMSE) of DOA estimation and the computational time with respect to the grid interval $r$ and SNR.
Assume the ULA composed of $M =7$ sensors with $d=\lambda/2$ is used to receive $K=2$
uncorrelated signals. The two signals uniformly come from intervals $[-30^\circ,-20^\circ]$ and $[0^\circ,10^\circ]$, respectively.
Assume that  the number of snapshots $T=30$, $\eta=2$, SNR $=$ $10$ and $0$~dB, and $r$ $=$ $1^\circ,2^\circ,4^\circ,6^\circ,8^\circ$ and $10^\circ$.  Fig. 1 shows the RMSE of DOA estimation versus grid interval based on 200 Monte Carlo runs.
It is seen that our method outperforms the state-of-the-art methods, especially when the grid interval is large. The reason is that the linear approximation adopted in  OGSBI will lead to a high modeling error when a coarse grid is used [see (\ref{problem2})]; while
our method can properly handle the modeling error through viewing the sampled locations in the coarse grid as the adjustable parameters.

Fig. 2 shows the total CPU time versus grid interval based on 200 Monte Carlo runs.
As can be seen from the figure, the computational times required by all the methods
decrease as the grid gets coarser. Our method is much faster than the other methods, especially when the grid interval is large. Therefore, the results in Figs. 1 and 2 recommend to use a coarser grid with $r= 4^\circ$ or $r=6^\circ$ , as it can give a fast DOA estimation but remain a reasonable accuracy.

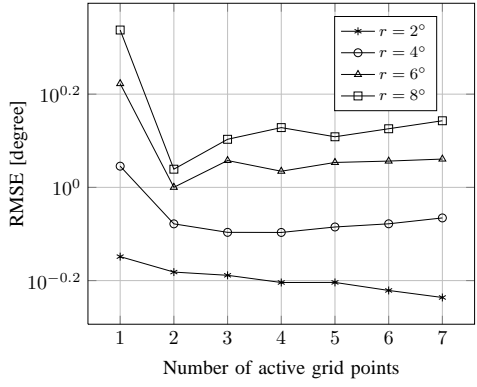
\begin{figure}
\center
\begin{tikzpicture}[scale=0.75]
\begin{semilogyaxis}[xlabel={Number of active grid points},
ylabel={RMSE [degree]},grid=major,
legend style={at={(0.77,0.96),font=\footnotesize},
anchor=north,legend columns=1}, ]
\addplot[mark=asterisk,black]  coordinates{
  (    1.0000,    0.7104)
  (    2.0000,    0.6581)
  (    3.0000,    0.6477)
  (    4.0000,    0.6251)
  (    5.0000,    0.6253)
  (    6.0000,    0.6010)
  (    7.0000,    0.5805)
};
\addplot[mark=o,black]  coordinates{
  (    1.0000,    1.1104)
  (    2.0000,    0.8351)
  (    3.0000,    0.8008)
  (    4.0000,    0.8004)
  (    5.0000,    0.8223)
  (    6.0000,    0.8353)
  (    7.0000,    0.8599)
};
\addplot[mark=triangle,black]  coordinates{
  (    1.0000,    1.6695)
  (    2.0000,    1.0003)
  (    3.0000,    1.1418)
  (    4.0000,    1.0830)
  (    5.0000,    1.1310)
  (    6.0000,    1.1387)
  (    7.0000,    1.1501)
};
\addplot[mark=square,black] coordinates{
  (    1.0000,    2.1756)
  (    2.0000,    1.0940)
  (    3.0000,    1.2681)
  (    4.0000,    1.3435)
  (    5.0000,    1.2839)
  (    6.0000,    1.3359)
  (    7.0000,    1.3897)
};
\legend{$r=2^\circ$, $r=4^\circ$, $r=6^\circ$ , $r=8^\circ$}
\end{semilogyaxis}
\end{tikzpicture}
\caption{RMSE of DOA estimate versus the number of active grid points with SNR $=$ $0$ dB.
 }\label{fig3}
\end{figure}

The last simulation investigates the effect of active grid points
on the DOA estimation performance.
Consider the same scenario as in Simulation 1, except for SNR $=$ $0$~dB.
Fig.~3 shows the RMSE of DOA estimation versus the number of active grid points based on 200 Monte Carlo runs.
Fig.~4 shows the total CPU time versus the number of active grid points.
It is shown that the choice of the number of  active grid points does not affect the performance much, as long as it is larger than the number of  signals $K$; however,  it does bring a light computational
cost, if a small value is chosen. Another observation is that if the grid is very coarse, knowing the exact value of $K$ can improve the DOA estimation performance.

%

\begin{figure}
\center
\begin{tikzpicture}[scale=0.75]
\begin{semilogyaxis}[xlabel={Number of active grid points},
ylabel={Time [second]},grid=major,
legend style={at={(0.67,0.86),font=\footnotesize},
anchor=north,legend columns=1}
]
\addplot[mark=asterisk,black]  coordinates{
  (    1.0000,   72.5805)
  (    2.0000,   80.2269)
  (    3.0000,   84.8572)
  (    4.0000,   92.3237)
  (    5.0000,  102.0196)
  (    6.0000,  108.9156)
  (    7.0000,  113.5554)
};
\addplot[mark=o,black]  coordinates{
  (    1.0000,    14.9155)
  (    2.0000,    20.6258)
  (    3.0000,    22.5570)
  (    4.0000,    24.9009)
  (    5.0000,    29.4173)
  (    6.0000,    30.4876)
  (    7.0000,    37.1329)
};
\addplot[mark=triangle,black]  coordinates{
  (    1.0000,    6.6464)
  (    2.0000,    9.2072)
  (    3.0000,   10.1548)
  (    4.0000,   12.7731)
  (    5.0000,   13.5306)
  (    6.0000,   15.9475)
  (    7.0000,   16.2000)
};
\addplot[mark=square,black] coordinates{
  (    1.0000,    5.7786)
  (    2.0000,    8.0707)
  (    3.0000,    8.3709)
  (    4.0000,   10.3360)
  (    5.0000,   11.3122)
  (    6.0000,   12.9172)
  (    7.0000,   13.5435)
};
\legend{$r=2^\circ$, $r=4^\circ$, $r=6^\circ$ , $r=8^\circ$ }
\end{semilogyaxis}
\end{tikzpicture}
\caption{Computational time versus the number of active grid points with SNR $=$ $0$ dB.}
\label{fig4}
\end{figure}
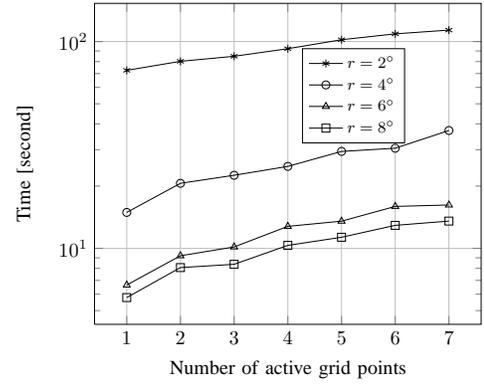

\section{Conclusion}
We have proposed a computationally efficient root SBL method that approaches the problem of DOA estimation with off-grid model error. Unlike the existing off-grid SBL method in \cite{yang2013off} that applies a linear approximation to the true DOA, our new method considers the sampled locations in the coarse grid as the adjustable parameters, and utilizes an EM algorithm to iteratively refine the grid. We further illustrate that each updated grid point can be simply achieved by the root of the polynomial (\ref{eqpolynomial}).
Simulation results demonstrate that the computational complexity is
significantly reduced and the modeling error can be almost
eliminated.


\bibliographystyle{IEEEtran}
\bibliography{rootSBL}

\end{document}